\documentclass[aps,showpacs,twocolumn,
superscriptaddress]{revtex4}
\usepackage{graphicx}
\usepackage{dcolumn}
\usepackage{bm}
\usepackage{color}
\usepackage[normalem]{ulem} 
\newcommand{\jk}[1]{{\textcolor{blue}{{#1}}}}

\usepackage[dvipsnames]{xcolor} 
\usepackage{hyperref}
\hypersetup{
  colorlinks=true,        
  linkcolor=blue,         
  citecolor=cyan,         
}
\usepackage{mathrsfs}
\usepackage[utf8]{inputenc}
\usepackage{mathtools}
\usepackage{doi}
\usepackage{amsmath}
\usepackage{amssymb}
\usepackage{multirow}

\begin{document}

\title{Parameter Constraints on Traversable Wormholes within Beyond Horndeski Theories through Quasi-Periodic Oscillations
}

\author{Farukh Abdulkhamidov} \email{farrux@astrin.uz}
\affiliation{School of Mathematics and Natural Sciences, New Uzbekistan University, Movarounnahr Str. 1, Tashkent 100000, Uzbekistan}
\affiliation{Research Centre for Theoretical Physics and Astrophysics, Institute of Physics, Silesian University in Opava, Bezru\v covo n\' am. 13, CZ-74601 Opava, Czech Republic}
 
\author{Petya Nedkova}
\email{pnedkova@phys.uni-sofia.bg}
\affiliation{Faculty of Physics, Sofia University,
    5 James Bourchier Boulevard, Sofia 1164, Bulgaria}

\author{Javlon Rayimbaev}
\email{javlon@astrin.uz}
\affiliation{Institute of Fundamental and Applied Research, National Research University TIIAME, Kori Niyoziy 39, Tashkent 100000, Uzbekistan}
\affiliation{Tashkent University of Applied Sciences, Gavhar Str. 1, Tashkent 100149, Uzbekistan}
\affiliation{Tashkent State Technical University, Tashkent 100095, Uzbekistan}

\author{Jutta Kunz}
\email{jutta.kunz@uni-oldenburg.de}
\affiliation{Institute of Physics, University of Oldenburg, D-26111 Oldenburg, Germany}

\author{Bobomurat Ahmedov}
\email{ahmedov@astrin.uz}
\affiliation{Institute of Theoretical Physics, National University of Uzbekistan, Tashkent 100174, Uzbekistan}
\affiliation{Ulugh Beg Astronomical Institute, Astronomy Str. 33, Tashkent 100052, Uzbekistan}


%

\date{\today}

\begin{abstract}
{\it Hunting} compact astrophysical objects such as black holes and wormholes, as well as testing gravity theories, are important issues in relativistic astrophysics. In this sense, theoretical and observational studies of quasiperiodic oscillations (QPOs) observed in (micro)quasars become helpful in exploring their central object, which can be a black hole or a wormhole. In the present work, we study the throat properties of traversable wormholes beyond Horndeski theory. Also, we investigate the circular motion of test particles orbiting the wormhole. We analyze the test particle's effective potential and angular momentum for circular orbits. Frequencies of radial and vertical oscillations of the particles around stable circular orbits have also been studied and applied in explaining the quasiperiodic oscillations mechanism in the relativistic precession (RP) model.  Finally, we obtain constraint values for the parameters of Horndeski gravity and the mass of the wormhole candidates using QPOs observed in the microquasars GRO J1655-40, GRS 1915+105 \& XTE J1550-564 and at the center of Milky Way galaxy through Monte-Carlo-Markovian-Chain (MCMC) analyses.

\end{abstract}

\maketitle

\section{Introduction}

While General Relativity (GR) has been well tested in both weak and strong gravity regimes and is in accordance with all current observations \cite{Will:2005va,Will:2018bme}, it also leaves us with several challenging open problems associated, in particular, with its quantization and the dark sector in cosmology.
This has caused a flurry of interest in alternative theories of gravity in recent years (see e.g., \cite{Berti:2015itd,CANTATA:2021ktz}).

Certainly, any viable alternative theory of gravity must adhere to solar system constraints.
However, in the strong gravity sector, current observational data are still less accurate, calling for future observations to test new theoretical predictions.
In our current multi-messenger era much insight into alternative theories of gravity can be gained from (future) gravitational wave observations as well as from the observations in the full electromagnetic spectrum and particle detections \cite{LIGOScientific:2017ync}.

Among the numerous alternative theories of gravity Horndeski theories \cite{Horndeski:1974wa,Deffayet:2013lga} have been very attractive, since they involve only additional scalar degrees of freedom, leading to a set of second-order generalized Einstein-scalar field equations. 
Thus these theories are free of Ostrogradski instabilities. 
Indeed, Horndeski theories provide interesting predictions for black holes and neutron stars \cite{Berti:2015itd} and also for cosmology \cite{CANTATA:2021ktz}.

A new interesting class of alternative theories of gravity that extends the Horndeski theories was introduced in \cite{Gleyzes:2014dya}.
Whereas their equations are higher order in derivatives, the propagating degrees of freedom still possess second-order equations, making them free from Ostrogradski instabilities, as well.
Such beyond Horndeski \cite{Gleyzes:2014dya} and DHOST theories \cite{Langlois:2015cwa,Crisostomi:2016czh} are based on ``degenerate” Lagrangians, whose kinetic matrix cannot be inverted, and the ensuing constraints result in a reduced number of physical degrees of freedom.
Interestingly, when starting from Horndeski theories, beyond Horndeski theories can be generated with certain types of transformations 
\cite{Gleyzes:2014qga,Crisostomi:2016tcp,Crisostomi:2016czh}.

A well-motivated set of Horndeski theories are the Einstein-scalar-Gauss-Bonnet gravities. 
Such theories arise, for instance, in the low-energy regime of string theory \cite{Gross:1986mw,Metsaev:1987zx}.
Although they always involve a scalar field, this field can be coupled in a variety of ways, yielding rather different phenomenology when applied to compact objects and cosmology.
This allows us to obtain constraints on the respective coupling constants.

One of the most interesting predictions of the gravitational theories is wormholes. In general relativity, they cannot exist without violating the energy conditions. However, in some modified theories of gravity such as Gauss-Bonnet gravity or the f(R) theories the wormhole throat can be supported by the gravitational interaction itself without requiring the presence of exotic matter \cite{Hochberg:1990,Fukutaka:1989,Ghoroku:1992, Lobo:2009, Kanti:2011, Kanti:2011yv, Antoniou:2019, Ibadov:2020}. In these theories, wormholes are viable alternatives of black holes. They could play a significant astrophysical role if their existence is confirmed observationally.

Searching for macroscopic wormholes alongside other exotic compact objects is currently one of the major goals of the astrophysical missions testing gravity in the strong field regime. Therefore, many theoretical works focus on studying their phenomenology. Studies of the quasinormal modes in wormhole geometries include \cite{Blazquez-Salcedo:2018, Azad:2022, Gonzalez:2022, Churilova:2021}. In the electromagnetic spectrum phenomenological effects arise primarily due to the specifics of the gravitational lensing in the wormhole spacetime \cite{Nedkova:2013, Gyulchev:2018, Bambi:2013, Tsukamoto:2012, Wielgus:2020, Huang:2023} and the properties of the accretion process in their vicinity \cite{Chakraborty:2016, Harko:2009, Zhou:2016}. Thus, it was demonstrated that wormholes can lead to a non-trivial morphology of the observable image of the accretion disk \cite{Paul:2019, Vincent:2020}. In addition to the primary disk image, they produce a series of bright rings at its center that proved to represent an observational signature of a rather general class of horizonless compact objects \cite{Gyulchev:2020, Gyulchev:2021, Eichhorn:2022}. Polarized emission from the accretion disk around the wormholes may further possess a distinctive twist of the polarization vector around the image \cite{Delijski:2022}. Thus, provided that we can observe radiation reaching through the throat of the wormhole, the properties of its linear polarization can serve as a distinctive signature for detecting traversable wormholes.

A possible way to test the spacetime of compact gravitational objects such as black holes and wormholes is by using spectrometrical analyses of radiation of accreting matter around them. However, the gravity of black holes in binary systems plays a crucial role in deriving all the radiation processes in the surrounding accretion disk. Astrophysical phenomena known as QPOs are found using Fourier analyses of the noisy continuous X-ray data from the accretion disk in (micro)quasars (containing black holes or neutron stars and companion stars as binary systems) \cite{Stella1998ApJL,Stella1999ApJ}, and they are classified as high frequency (HF) when the peak frequencies are about 0.1 to 1 kHz and low frequency (LF) when the frequencies are less than about 0.1 kHz \cite{Stella1999ApJ}.
Although high-accuracy measurements of QPO frequencies have been observed in binary systems, no exact and unique physical mechanisms for QPOs have been found yet. This problem is currently under active discussion to test gravity theories and measure the inner edge of the accretion disc. Such measurements may provide valuable information about ISCO radii and black hole parameters. In our previous studies \cite{Rayimbaev2021Galax...9...75R,Rayimbaev2021QPO,Rayimbaev2022CQGra,Rayimbaev2022EPJCEMSQPO,Rayimbaev2022IJMPDQPOcharged,Rayimbaev2022PDU,2023Galax..11...95R,2023Univ....9..391M,Rayimbaev2023AnPhy.45469335R,Rayimbaev2023EPJC...83..572R,Rayimbaev2023EPJC...83..730Q}, we have shown that studies of the QPO orbit may help to determine the ISCO radius which lies near the orbit since the distance between these orbits is in the order of the error of measurements using the relativistic precision model.

The quasi-periodic oscillations in wormhole spacetimes were previously studied within the geodesic models in \cite{Deligianni:2021, Deligianni:2021hwt, DeFalco:2021, Stuchlik:2021, Stuchlik:2021a, DeFalco:2023}.
Using the resonance models, it was demonstrated that twin peak oscillations can be described by a more diverse resonance structure compared to black holes \cite{Deligianni:2021, Deligianni:2021hwt}. Wormhole geometries may allow for parametric and forced resonances of lower order which produce more significant observable signals. In addition, resonances may be excited in an arbitrarily close vicinity of the wormhole throat, further amplifying the signals and allowing probes to be placed deep inside the gravitational field of the compact object.

The possibility to explain the observed quasi-periodic oscillations of astrophysical X-ray sources through an underlying wormhole spacetime was investigated in \cite{Stuchlik:2021}. Using the Simpson-Visser metric, which interpolates between different compact object geometries, including wormholes, the QPOs were modeled as parametric resonances between the epicyclic frequencies. The results were used to fit the observational data for microquasars and active galactic nuclei with available independent mass measurements.

The aim of this paper is to investigate further the compatibility of the wormhole geometry with the properties of certain X-ray sources using QPO measurements. We explore a traversable wormhole spacetime that arises as an exact solution within a beyond Horndeski gravitational theory \cite{Bakopoulos:2021liw} and interpret the twin peak frequencies applying the geodesic precession model. Using a Markov Chain Monte Carlo algorithm we fit the theoretical predictions for the QPO frequencies to the observational data for the microquasars GRO J1655-40, GRS 1915+105, and XTE J1550-564  and the galactic target Sgr A*. In this way, we provide constraints on the parameters of the beyond Horndeski gravitational theory and the possible wormhole geometry.

The paper is organized as follows. In Section II we review the wormhole solution in the beyond Horndeski theory, which we explore. In sections III and IV we study the circular timelike geodesics in the equatorial plane and derive the epicyclic frequencies. In Section V we describe the QPO frequencies using the geodesic precession model investigating their properties. In section VI we perform a fit of the observational data for the microquasars GRO J1655-40, GRS 1915+105, and XTE J1550-564  and the center of our galaxy Sgr A* constraining the parameters of the gravitational theory as well as the wormhole mass and the QPOs excitation radius. In the last section, we present our conclusions.

\section{Beyond Horndeski Wormholes}

The traversable beyond Horndeski wormhole solutions obtained by 
Bakopoulos et al. \cite{Bakopoulos:2021liw}
are based on the black hole seed solution
of Lu-Pang \cite{Lu:2020iav,Hennigar:2020lsl,Fernandes:2020nbq}
in a Horndeski theory defined by the functions 
\begin{eqnarray*}
G_2&=&8\alpha \bar{X}^2,\\
G_3&=&-8\alpha \bar{X},\\
G_4&=&1+4\alpha \bar{X},\\ 
G_5&=&-4\alpha \ln\bar{|X|},
\end{eqnarray*}
where
$\bar{X}=-\frac{1}{2}\,\partial_\mu\bar{\phi}\,\partial^\mu\bar{\phi}$, $\bar{\phi}$ represents the Horndeski scalar field, and $\alpha$ is the Horndeski coupling constant, $\alpha \geq 0$ \cite{Clifton:2020xhc,Charmousis:2021npl,Bakopoulos:2021liw}.

This black hole seed solution is static and spherically symmetric with metric parametrization
\begin{equation} ds^2=g_{tt}dt^2+g_{rr}dr^2+g_{\theta \theta}d\theta^2+g_{\phi \phi}d\phi^2\ , 
\end{equation}
where $g_{tt}=-\bar{h}(r), \ g_{rr}=1/\bar{f}(r), \ g_{\theta \theta}=r^2,$ and $g_{\phi \phi}=r^2\sin^2\theta$. 
The metric functions and scalar field function $\bar{\phi}(r)$ are \cite{Lu:2020iav} 
\begin{eqnarray}
   \bar{h}(r)&=&\bar{f}(r)= 1+\frac{r^2}{2 \alpha} 
   \left(1\ - \sqrt{1+\frac{8\alpha M}{r^3}}\right), 
   \label{seed_h_f}\\  
   \bar\phi'&=&\frac{\sqrt{\bar{h}}- 1}{r \sqrt{\bar{h}}},
   \label{seed_phi} 
\end{eqnarray}
respectively, and the prime indicates a partial derivative with respect to $r$. Note, that this corresponds to the asymptotically flat branch of the seed solution \cite{Lu:2020iav}.
The parameter $M$ then corresponds to the mass of the seed black hole, and the radius of the outer horizon is given by
$r_+= M + \sqrt{M^2 -\alpha}$.
For $\alpha > M^2$, the seed solution becomes a naked singularity.

The Horndeski seed solution is turned into a beyond Horndeski solution by a disformal transformation of the metric (see, e.g., \cite{Zumalacarregui:2013pma,Crisostomi:2016tcp,BenAchour:2016cay,BenAchour:2016fzp,Bakopoulos:2021liw})
\begin{equation}
\label{dis_trans}
g_{\mu \nu}=\bar{g}_{\mu \nu}-D(\bar{X})\; \nabla_\mu \bar \phi \,\nabla_\nu \bar \phi\,,
\end{equation}
where the new beyond Horndeski metric $g_{\mu \nu}$ is obtained from the seed Horndeski metric $\bar{g}_{\mu \nu}$ by making use of a \textit{disformability} function $D(\bar X)$ leading to\cite{Bakopoulos:2021liw}
\begin{eqnarray} 
\phi &=&\bar{\phi}, \label{new_p}\\
h&=&\bar{h}, \label{new_h}\\ 
f&=&\frac{\bar{f}}{1+2 D \bar{X}}
  = h W(\bar X)^{-1}, \label{new_f}\\
X&=&\frac{\bar{X}}{1+2 D \bar{X}}. 
\label{new_X}
\end{eqnarray}

As shown by Bakopoulos et al.~\cite{Bakopoulos:2021liw} the crucial transformation function $W(\bar X)^{-1}$ needs to satisfy certain constraints to obtain a traversable wormhole solution from the seed.
In particular, the radial coordinate at the throat $r_0$, where $W(\bar X)^{-1}$ has a root, should be chosen such that $r_0>r_+$, and $W(\bar X)^{-1} > 0$ for $r>r_0$.
Moreover, $W(\bar X)^{-1} \to 1$ for $r\to \infty$ to preserve asymptotic flatness.
These constraints led to the choice \cite{Bakopoulos:2021liw}
\begin{eqnarray}
W(\bar X)^{-1}&=&1-\frac{r_0}{\lambda}\,\sqrt{-2\bar X}, 
\nonumber\\
    &=& 1-\frac{r_0}{\lambda  r}\left( 1-\sqrt{h}  \right),
    \label{W_inv}
\end{eqnarray}
since
\begin{equation}
    \bar X = - \frac{1}{2}\,\bar h \bar{\phi}'^2
    = -\frac{1}{2}\,\frac{(\sqrt{\bar{h}}-1)^2}{r^2},
\end{equation}
and $\lambda$ is a new dimensionless positive parameter.

The three constants $\alpha$, $\lambda$, and $M$ entering the wormhole solutions may not all be chosen independently, but they have to satisfy constraints.
The physical requirement, that $r_0$ be the wormhole throat translates to $W(\bar X(r_0))^{-1}=0$.
Eq.~(\ref{W_inv}) then yields
\begin{equation}
    h(r_0)=(1-\lambda)^2.
    \label{r_0_lambda}
\end{equation}
Moreover, it is required that
\begin{equation}
0 < \lambda < 1
\end{equation}
since for $\lambda=0$ $r_0$ would move to infinity and for $\lambda=1$, $r_0=r_+$ \cite{Bakopoulos:2021liw}.
Inserting the metric function $h$ in eq.~(\ref{r_0_lambda}) and solving for the throat radius $r_0$ one obtains the relation
\begin{equation}
    r_0=\frac{ M + \sqrt{M^2- \alpha \lambda^3\, (2-\lambda)^3}}{\lambda(2-\lambda)},
    \label{r_0_M_lambda}
\end{equation}
and since the radicand in eq.~(\ref{r_0_M_lambda}) should not become negative, this leads to the bound \cite{Bakopoulos:2021liw}
\begin{equation}
    \alpha\leq\frac{M^2}{\lambda^3\,(2-\lambda)^3}. \label{alpha_M_lambda}
\end{equation}
Also, we will take into account observational constraints on the Horndeski coupling constant $\alpha$ obtained, in particular, from inflation in the early universe and binary systems with black holes \cite{Clifton:2020xhc,Charmousis:2021npl}.

\begin{figure}
    \centering
    \includegraphics[width=0.45\textwidth]{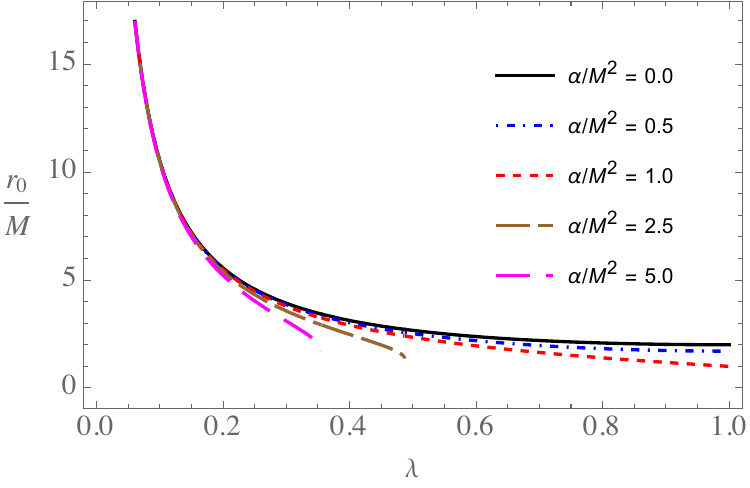}
    \includegraphics[width=0.45\textwidth]{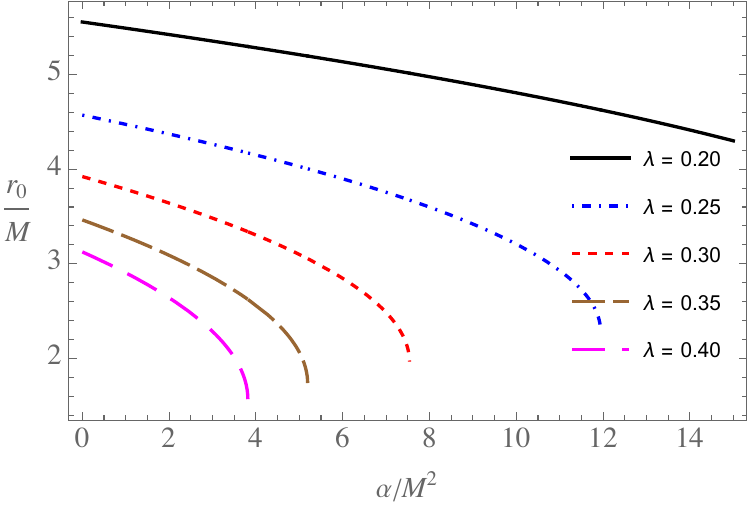}
    \includegraphics[width=0.45\textwidth]{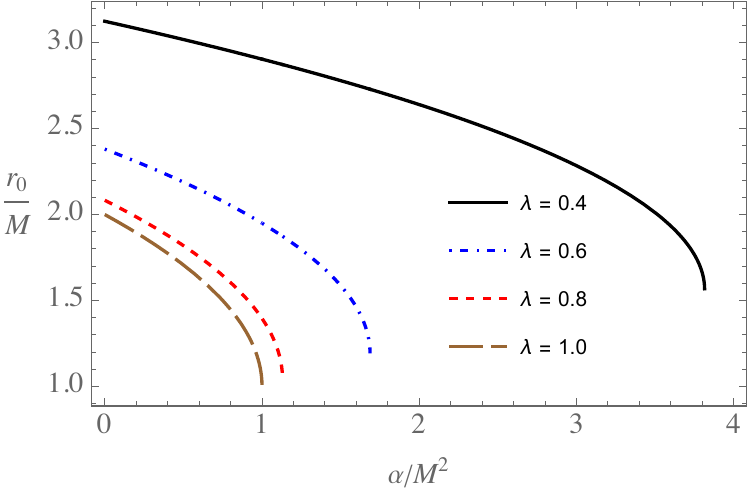}
    \caption{Dependence of the throat of the wormhole on the  parameters $\lambda$ and $\alpha$.}
    \label{fig:throat}
\end{figure}

The behavior of the wormhole throat as a function of the solution parameters is presented in Fig. $\ref{fig:throat}$. We see that its location moves to smaller values of the radial coordinate when any of the parameters $\alpha$ or $\lambda$ increases.

We should note that the radial coordinate $r$ does not cover the wormhole spacetime completely. By definition it ranges from the wormhole throat, where we have a coordinate singularity, to $r\rightarrow +\infty$ representing one of the asymptotic ends of the spacetime. In order to define a global coordinate system that also covers the second asymptotic end, we should extend the radial coordinate through the wormhole throat. This is possible for example by defining the coordinate  $\ell=\pm \sqrt{r^2-r_0^2}$. It describes two identical spacetime regions glued at the throat of the wormhole $r=r_0$, taking negative or positive values in each of them. The asymptotic ends correspond to $\ell\rightarrow\pm\infty$, respectively, and the wormhole throat is located at $\ell=0$.

\section{Circular motion around a traversable wormhole in beyond Horndeski theory\label{partdynam}}

In this section, we first investigate the circular motion of electrically neutral test particles in the spacetime of a traversable wormhole in beyond Horndeski theory, together with the energy and angular momentum of test particles in stable circular orbits. 

\subsection{Equations of motion}

First, we start deriving the equations of motion using the following Lagrangian for test particles, 
\begin{eqnarray}
L_{\rm p}=\frac{1}{2}m g_{\mu\nu} \dot{x}^{\mu} \dot{x}^{\nu} \ , 
\end{eqnarray}
where $m$ is the mass of the particle. 
We obtain the integrals of motion such as the specific energy ${\cal E}=E/m$, and angular momentum ${\cal L}=L/m$ of the particles in the following form
\begin{eqnarray}
\label{consts1}
 g_{tt}\dot{t}=-{\cal E}\ , \qquad g_{\phi \phi}\dot{\phi} = {\cal L}. 
\end{eqnarray}
Spherical symmetry allows us to set $\theta=\pi/2$ and $\dot{\theta}=0$.
The equation for the radial motion of the test particles can then be obtained using the normalization condition, $g_{\mu \nu}u^{\mu}u^{\nu}=-1$ and taking into account the integrals of the motion (\ref{consts1})

\begin{equation}
 - g_{rr} g_{tt} \dot{r}^2={\cal E}^2
  + {g_{tt}} \left(1 + \frac{{\cal L}^2}{r^2}\right) \ .
\end{equation}
This leads to the effective potential \cite{Bakopoulos:2021liw}
\begin{equation}
\label{effpotentail}
     V_{\rm eff}(r) = h \left(1 + \frac{{\cal L}^2}{r^2}\right) \ .
   \end{equation}

\begin{figure}[ht!]
    \centering
    \includegraphics[width=0.45\textwidth]{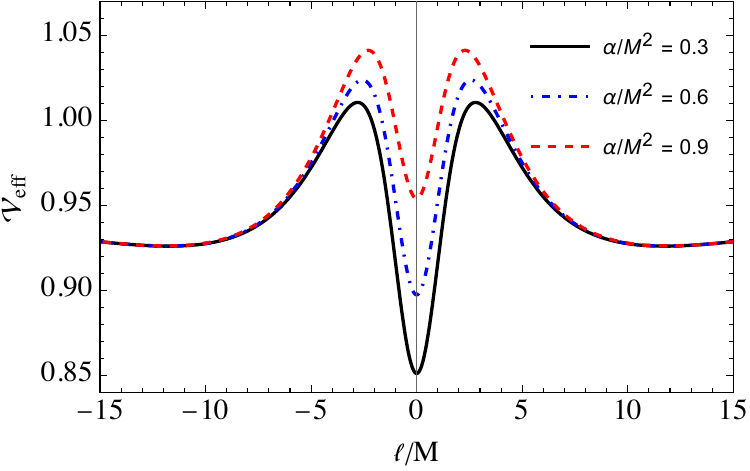}
    \includegraphics[width=0.45\textwidth]{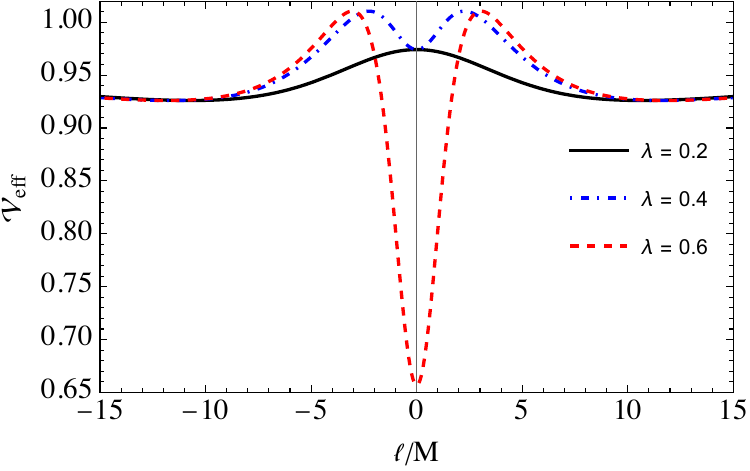}
    \caption{The dependence of the effective potential on the radial coordinate \jk{$\ell$} is illustrated for various values of the spacetime parameters. 
    The upper panel shows $\lambda$ fixed to 0.5 while $\alpha/M^2$ varies, while the lower panel presents $\alpha/M^2$ set to 0.3 with varying $\lambda$ values.}
    \label{fig:effpot}
\end{figure}

In Fig.~\ref{fig:effpot}, we show the dependence of the effective potential on the radial coordinate $\ell$, where $\ell=\pm \sqrt{r^2-r_0^2}$, for various values of $\alpha$ and $\lambda$.
In the top panel, we show the effective potential for $\lambda=0.5$ and several values of $\alpha$.
We find that an increase in $\alpha$ causes an increase in the effective potential for small values of the radial coordinate $\ell$. The gravitational effect of the parameter $\alpha$ becomes less pronounced for large values $|\ell|>5M$, leading only to small variations in the effective potential.

In the bottom panel, we illustrate the effect of $\lambda$ on the effective potential for the case $\alpha=0.3M^2$.  
We see that variation of the parameter $\lambda$ can lead to modification of the qualitative behavior of the effective potential. 
For larger values of $\lambda$ we observe that the effective potential possesses two maxima located symmetrically on both sides of the wormhole throat and a minimum at the throat. 
Decreasing the parameter $\lambda$, the two maxima approach the throat and at some critical value of $\lambda$ they reach it merging into a single maximum and causing the minimum of the effective potential to disintegrate. 
This transition has an impact on the qualitative behavior of the timelike geodesics in the wormhole spacetime. 
While the first configuration allows for bound particle orbits in the vicinity of the wormhole throat, the second one possesses only a single unstable circular orbit located at the throat.

Next, we explore the stable circular orbits of test particles around the traversable wormhole using the standard conditions
\begin{eqnarray} \label{conditions}
V_{\rm eff}={\cal E}^2, \qquad V_{\rm eff}'=0  \ .
\end{eqnarray}
The expressions for the angular momentum and the energy of test particles corresponding to circular orbits take the form
\begin{eqnarray} \label{lcre} 
{\cal L}^2=\frac{r^3 h'(r)}{2 h(r)-r h'(r)}, 
\quad
{\cal E}^2=\frac{2 h(r)^2}{2 h(r)-r h'(r)}.  
\end{eqnarray}

\begin{figure}
    \centering
    \includegraphics[width=0.4\textwidth]{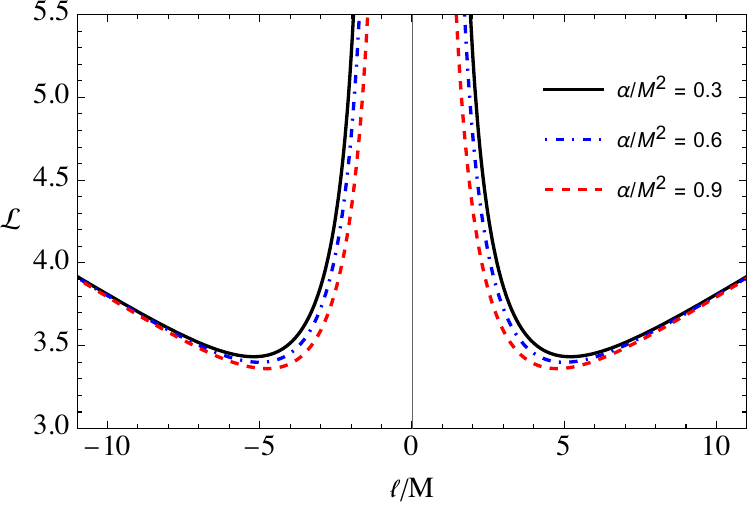}
    \includegraphics[width=0.4\textwidth]{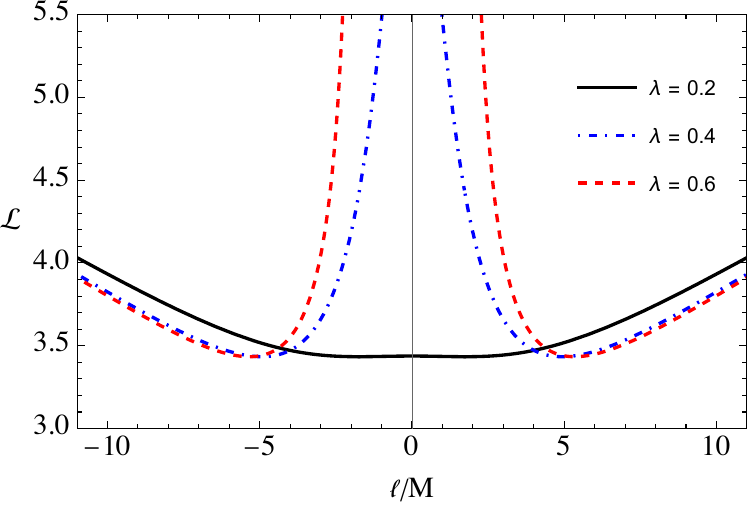}
    \caption{
    The dependence of the angular momentum of circular orbits on the radial coordinate $\ell$ is presented.
    In the top panel, the dependence is shown for varying $\alpha/M^2$ values, with $\lambda$ held constant at 0.5. Conversely, the lower panel illustrates the angular momentum 
    for several $\lambda$ values, while maintaining $\alpha/M^2$ at 0.3.}
    \label{fig:EL}
\end{figure}

The dependence of the angular momentum on the radial coordinate $\ell$ is shown in Fig.~\ref{fig:EL} for different values of the parameters $\alpha$ and $\lambda$.  
The figure shows that the angular momentum decreases slightly with an increase of both the parameters $\alpha$ and $\lambda$. 
However, at distances less than about, $\ell \simeq 4 M$ the angular momentum increases as $\lambda$ increases. 
Also, the angular momentum at $\ell=0$ increases as $\lambda$ increases.

\section{Fundamental frequencies of test particles~\label{fundfreq}}

Fundamental frequencies around astrophysical compact gravitational objects, such as black holes and wormholes, are related to the motion of matter or particles in the spacetime of the objects in the strong gravitational field regime. These frequencies can be used to probe the properties of black holes and wormholes, such as their mass, spin, and gravity parameters.
Now we investigate fundamental frequencies of test particles orbiting a traversable wormhole in beyond Horndeski theory. We start by calculating the angular velocity of the test particles in Keplerian orbits and the frequencies of the radial and vertical oscillations of the particles around the stable circular orbits.

The angular velocity of the test particles around central compact gravitating objects in circular orbits is called Keplerian frequency and is defined as $\Omega_K=\dot{\phi}/\dot{t}$ where the dot denotes differentiation with respect to the affine parameter along the geodesic.
The expression for the Keplerian frequency in the spacetime around a traversable wormhole in beyond Horndeski theory takes the following form
\begin{equation}\label{omegaKep}
\Omega_K^2={\frac{3 \text{M}}{r^{3} \sqrt{1+\frac{8 \alpha  \text{M}}{r^{3} }}}-\frac{\sqrt{1+\frac{8 \alpha  \text{M}}{r^{3} }}-1}{2 \alpha }}.
\end{equation}

The harmonic oscillations of the test particles around the central massive object are a result of the balance between the gravitational force of the object and the inertia of the test particle.

In fact, when a test particle is placed on a stable circular orbit around the object, it is subject to a centripetal force that is directed towards the object. This force is balanced by the gravitational force of the object, which is directed towards the center of it. If the test particle is slightly perturbed from its circular orbit, such as $r_0+\delta r$ and $\pi/2+\delta \theta$, it will experience a restoring force that will tend to bring it back to its original orbit. This restoring force is a result of the gradient of the gravitational field. The restoring force can be resolved into two components: a radial component and a vertical component. The radial component of the restoring force is responsible for the radial oscillations of the test particle, while the vertical component of the restoring force is responsible for the vertical one.

Now, here, we assume test particles orbiting a traversable wormhole along stable circular orbits which oscillate along the radial and the vertical directions. 
The following equations can evaluate the frequencies of the radial and the vertical oscillations \cite{Bardeen1972ApJ}:  
\begin{eqnarray}
\frac{d^2\delta r}{dt^2}+\Omega_r^2 \delta r=0\ , \quad \frac{d^2\delta\theta}{dt^2}+\Omega_\theta^2 \delta\theta=0\ , 
\end{eqnarray}
where 
\begin{eqnarray}
&&\Omega_r^2=-\frac{1}{2g_{rr}(u^t)^2}\partial_r^2V_{\rm eff}(r,\theta)\Big |_{\theta=\pi/2}\ ,
\\
&&\Omega_\theta^2=-\frac{1}{2g_{\theta\theta}(u^t)^2}\partial_\theta^2V_{\rm eff}(r,\theta)\Big |_{\theta=\pi/2}\ ,
\end{eqnarray}
are the radial and vertical frequencies, respectively. 

After some simple algebraic calculations, one may easily obtain the expressions for the frequencies,
\begin{widetext}
\begin{eqnarray}\label{wr}
\nonumber
\frac{\Omega_r^2}{\Omega_K^2} &=& \Big[4 \alpha  M^2 \left(10 \alpha -3 \sqrt{8 \alpha  M r+r^4}+21 r^2\right)-4 r^{9/2} \sqrt{8 \alpha  M+r^3}+4 r^6 + \\ &+& M \left(-32 \alpha  r^{3/2} \sqrt{8 \alpha  M+r^3}-15 r^{7/2} \sqrt{8 \alpha  M+r^3}+15 r^5+46 \alpha  r^3\right)\Big]  \times \\ \nonumber &\times& \frac{\left(2 \lambda  r-2 r_0\left(\sqrt{\frac{2 \alpha -\sqrt{8 \alpha  M r+r^4}+r^2}{2\alpha }}-1\right)   \right)}{2 \lambda  r \left(8 \alpha  M+r^3\right) \left(2 \alpha  M+r^{3/2} \left(-\sqrt{8 \alpha  M+r^3}\right)+r^3\right)},
\\ \label{wt}
\Omega_\theta &=&\Omega_K= \Omega_\phi\ .
\end{eqnarray}
\end{widetext}

It is worth noting that, in our further analysis, we express all the frequencies in the unit of Hz, i.e. we obtain for the epicyclic frequencies $\nu_i = \frac{1}{2\pi}\frac{c^3}{GM} \Omega_i$, 
where $c=3\cdot 10^8 \rm m/sec$ is the speed of light in vacuum, and $G=6.67\cdot 10^{-11}\rm m^3/(kg^2\cdot sec)$ is the gravitational Newtonian constant.

\begin{figure*}[ht!]
    \centering
    \includegraphics[width=0.32\textwidth]{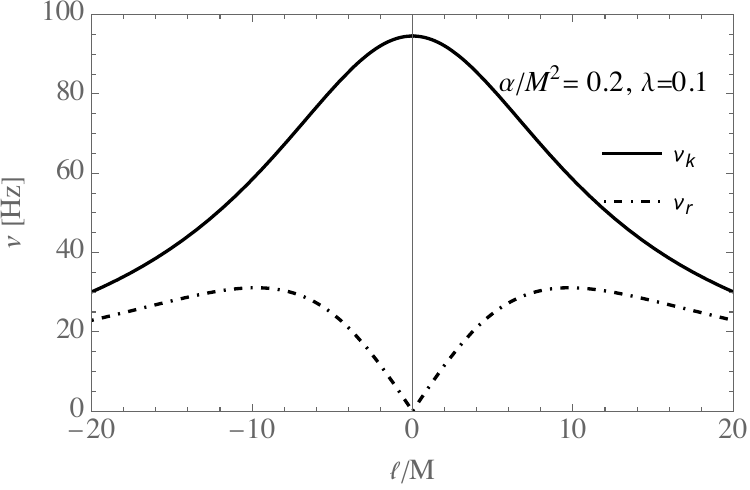}
    \includegraphics[width=0.32\textwidth]{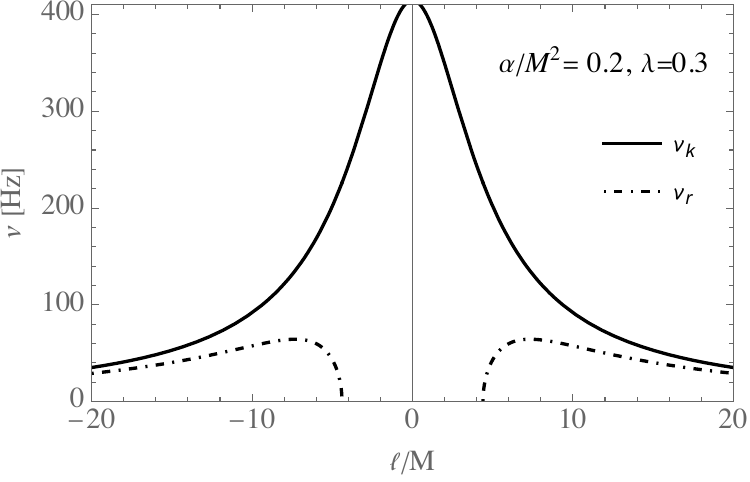}
    \includegraphics[width=0.32\textwidth]{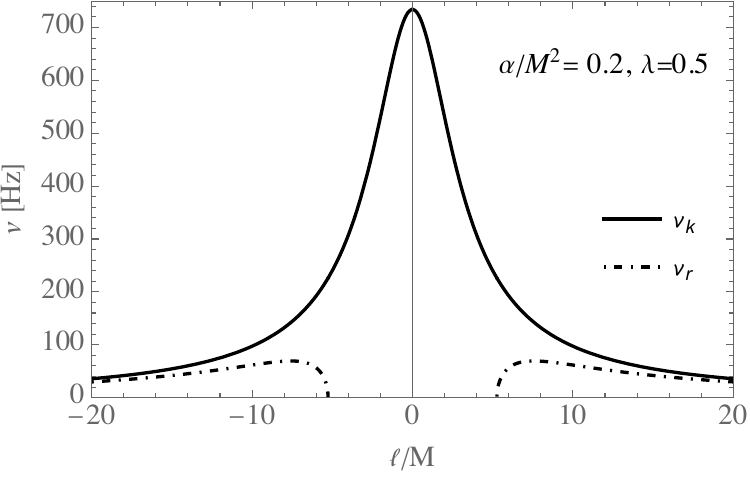}
    \includegraphics[width=0.32\textwidth]{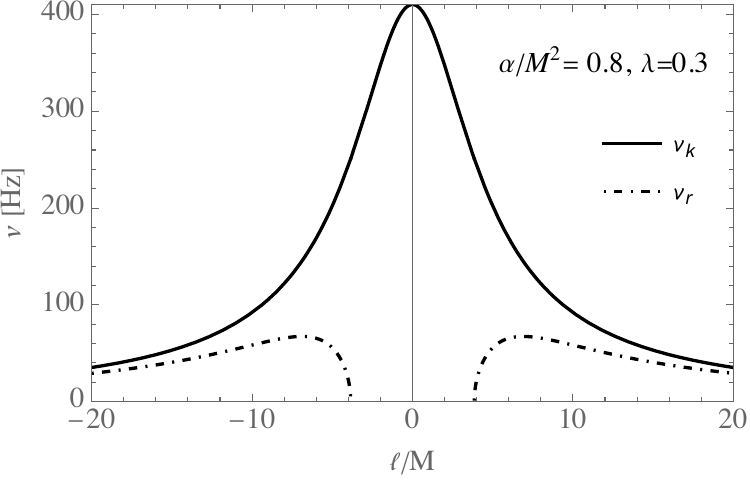}
    \includegraphics[width=0.32\textwidth]{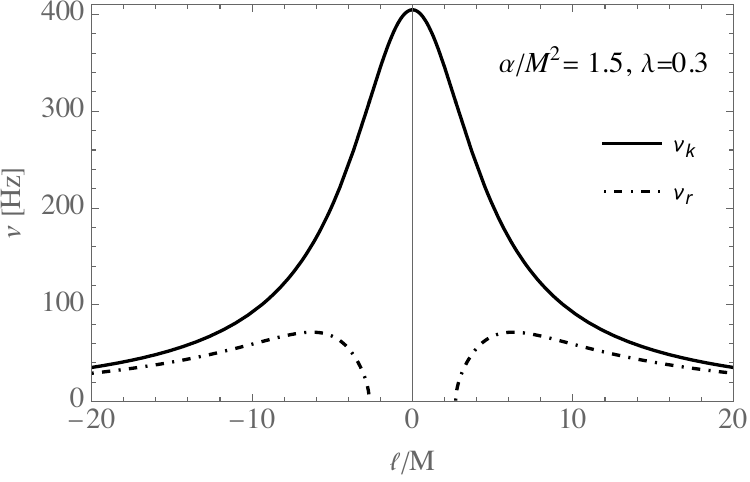}
    \includegraphics[width=0.32\textwidth]{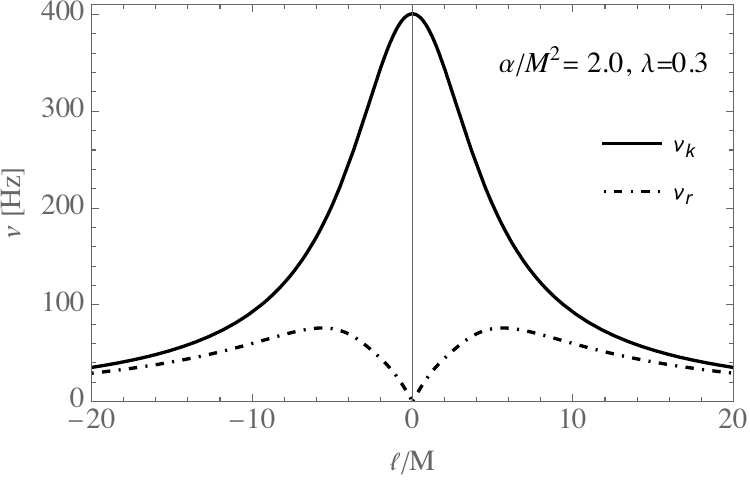}
    \caption{Radial dependence of the small oscillations $\nu_r$, and $\nu_k$ of neutral particles around a wormhole in beyond Horndeski theory having a mass $M = 10 M_{\odot}$. }
    \label{fig:nurnuK}
\end{figure*}

In Fig. \ref{fig:nurnuK} we demonstrate the behavior of the profiles of the Keplerian and the radial oscillation frequencies, when we vary the parameters $\alpha$ and $\lambda$. 
It is seen that when $\alpha=0.2M^2$ and $\lambda=0.1$, and $\alpha=2M^2$ and $\lambda=0.3$ the radial frequency goes to zero at $\ell=0$. In fact, the radial oscillations equal zero at the ISCO.
This means that the ISCO of the particles coincides with the throat of the wormhole. 
As the parameter $\lambda$ increases, the ISCO goes far from the wormhole throat. 

It is also observed that the effect of the coupling parameter $\alpha$ is visible near $\ell=0$, and an increase of $\alpha$ causes a slight decrease in the Keplerian frequency values. However, the parameter $\lambda$ increases it.

\section{Quasi-Periodic Oscillation Frequencies\label{qposec}}

Astrophysical compact objects, such as black holes and wormholes, do not emit electromagnetic radiation themselves. 
However, they do cause the curvature of spacetime, which in turn affects the motion of matter in their vicinity.
This motion can generate electromagnetic radiation, which can be observed in the form of an accretion disk. 

Quasi-periodic oscillations (QPOs) are a type of variability observed in the electromagnetic radiation from accretion disks.
QPOs are characterized by their frequency, which can be either high (HF QPOs) or low (LF QPOs). 
HF QPOs have frequencies in the range of 0.1 to 1 kHz, while LF QPOs have frequencies below 0.1 kHz. QPOs have been observed in the electromagnetic radiation from accretion disks around various astrophysical objects, including black holes, neutron stars, white dwarfs, and their binary systems ~\cite{Ingram2016MNRAS}. 
The origin of QPOs is not fully understood, but they are thought to be related to the dynamics of the accretion disk. 
The study of QPOs can provide insights into the physics of accretion disks and the properties of the compact objects they orbit. 
For example, the frequency of HF QPOs has been linked to the size of the inner region of the accretion disk, which is in turn related to the mass and spin of the compact object. 
Consequently, one can conclude that QPOs are a valuable tool for studying the physics of accretion disks and the properties of the compact objects they orbit.

QPOs observed in low-mass X-ray binaries (LMXBs) are thought to be produced by a mechanism different from that observed in other astrophysical objects. 
This is because LMXBs contain neutron stars that have strong magnetic fields. 
The magnetic field of a neutron star can interact with the accretion disk in a way that produces QPOs. 
The source of the electromagnetic radiation in the accretion disk is thought to be the oscillation of test-charged particles. 
These particles radiate with the same frequency as their oscillation frequency. 
The relativistic precession (RP) model explains the existence of QPOs due to the quasiharmonic oscillations of charged particles in the radial and angular directions around black holes and wormholes.

In the RP model, the upper frequency of a twin-peaked QPO is equal to the orbital frequency of the particle, while the lower frequency is equal to the difference between the orbital frequency and the radial oscillation frequency. 
In other words, the upper and lower frequencies of a QPO with twin peaks are given by: $\nu_U=\nu_\phi$ and $\nu_L=\nu_\phi-\nu_r$,  where, $\nu_U$ and $\nu_L$ are the upper and lower frequencies, respectively, and $\nu_\phi$ is the orbital frequency and $\nu_r$ is the radial oscillation frequency. 

\begin{figure}
    \centering
    \includegraphics[width=0.45\textwidth]{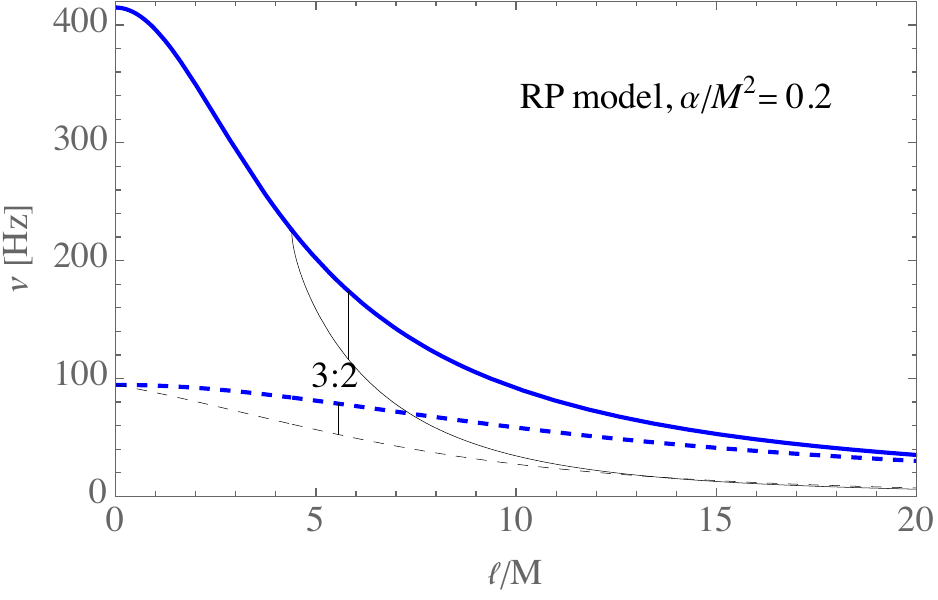}
    \includegraphics[width=0.45\textwidth]{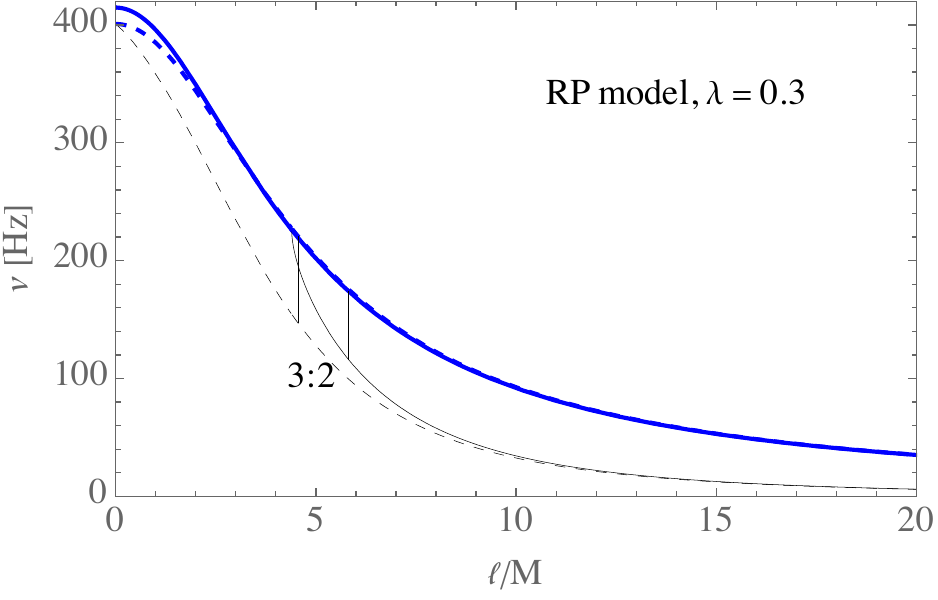}
    \caption{The radial profile of the upper (\(\nu_{U}\), blue) and lower (\(\nu_L\), black) frequencies is presented for the RP model. 
    The upper panel displays results for two distinct values of \(\lambda\) (solid lines for $\lambda=0.3$, while dashed lines for $\lambda = 0.1$) while maintaining \(\alpha/M^2\) at a constant. 
    Conversely, the lower panel showcases variations for two specific values of \(\alpha/M^2\) (solid lines for $\alpha/M^2=0.2$, while dashed lines for $\alpha/M^2 = 2.0$), with \(\lambda\) held constant. 
    Furthermore, $r_{3:2}$ resonance radii are also displayed by solid vertical lines.}
    \label{fig:upperlower1}
\end{figure}

Figure \ref{fig:upperlower1} presents the radial dependence of the upper and lower frequencies of twin QPOs with blue and black lines, respectively. 
In the upper panel, the parameter $\alpha$ is fixed as $\alpha=0.2M^2$ and the solid line corresponds to $\lambda=0.3$ and the dashed line to $\lambda=0.1$. 
It is observed from this panel that an increase in the $\lambda$ parameter causes a decrease in the upper and lower frequencies, and the QPO orbit, with the frequency ratio of 3:2, slightly shifts out. 
In the bottom panel, we specify the solid line for $\alpha=0.2M^2$ and the dashed one for $\alpha=2.0M^2$ cases by fixing the parameter $\lambda=0.3$. 
It is seen that an increase of $\alpha$ reduces $\nu_U$ slightly near the wormhole throat (up to about $\ell \sim 3M$), and, at far distances, its effect is invisible. 
However, the effects of the $\alpha$ parameter on the lower frequencies are sufficient.

\begin{figure*}
    \centering
    \includegraphics[width=0.42\textwidth]{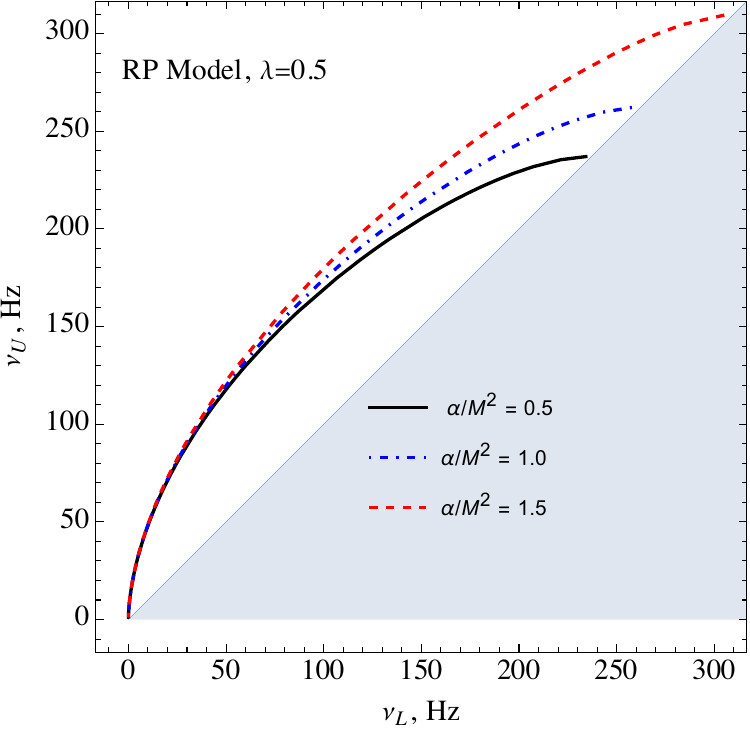}
    \includegraphics[width=0.42\textwidth]{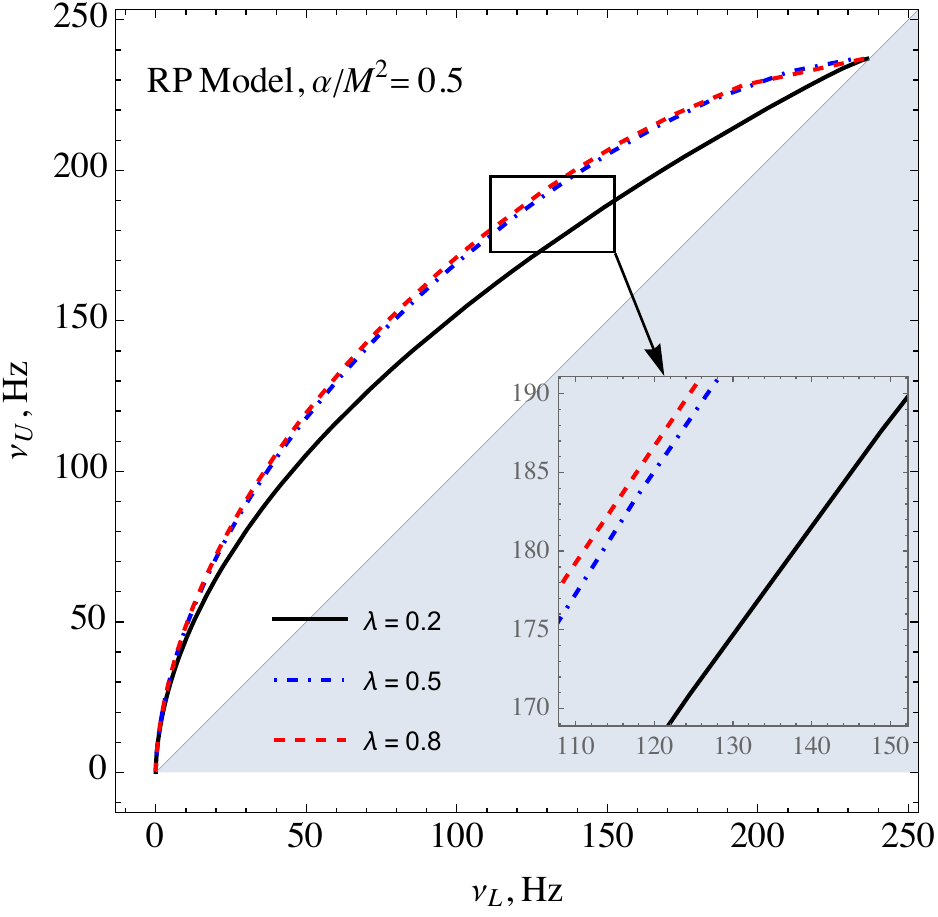}
    \caption{Relations between the frequencies of the upper and lower peaks of the twin-peak QPOs in the RP model around the wormhole having mass $M=10 M_\odot$}
    \label{fig:uplower}
\end{figure*}

We perform an analysis of the relationships between the upper and lower frequencies of the twin QPOs for different values of $\alpha$ and $\lambda$ in Fig.~\ref{fig:uplower}.  

In the left panel, we show the relationships for different values of $\alpha$ in $\lambda=0.5$, and in the right panel, the relationships are given for different values of $\lambda$ by fixing $\alpha=0.5M^2$. 
It is obtained that one can observe an increase in the highest values of upper and lower frequencies corresponding to those generated at ISCO, due to an increase in $\alpha$. 
However, the highest value does not change in the variation of $\lambda$ and it only depends on $\alpha$. 
From the right panel, it is also seen that the frequency ratio increases slightly with increasing $\lambda$.

\section{Constraints on the wormhole mass in beyond Horndeski theory \label{section5}}

In this section, we investigate four X-ray binary systems and endeavor to derive constraints on the parameters associated with our wormhole model within the beyond Horndeski theory framework. This will be achieved by analyzing the QPOs' data. Celestial objects of interest include $Sgr A^*$, GRO J1655-40, GRS 1915 + 10, and XTE 1550-564.
Finally, we shall present the best-fit values within the parameter space, which have been obtained through the utilization of a Markov Chain Monte Carlo (MCMC) code analysis.

\begin{table*} 
\renewcommand\arraystretch{1.5}
\caption{\label{binary}%
 The mass, upper, and lower frequencies of the QPOs from $Sgr A^*$
 and the X-ray binaries selected for analysis.}

\begin{tabular}{lcccc}
\hline\hline

 & $Sgr A^*$ & GRO J1655-40  & GRS 1915+105  & XTE J1550-564\\
  \hline
     $ M\; (M_{\odot})$ & $2.6 \div 4.4 \times 10^6$ \cite{2004ragt.meet....1A}  & 6.03$\div$6.57 \cite{Kolos:2020ykz} & $10.4 \div 14.2$ \cite{Remillard:2006fc}& 8.49$\div$ 9.71 \cite{Remillard:2002cy, Orosz:2011ki} \\
     
     $\nu_{\text{up}}$(Hz) & $1.445_{-0.16}^{+0.16}\times 10^{-3}$ \cite{2004ragt.meet....1A} & 451$\pm$ 5 \cite{Kolos:2020ykz}  & 168 $\pm$ 3 
 \cite{Remillard:2006fc}  & 276$\pm$ 3 \cite{Remillard:2002cy}  \\
     
     $\nu_{\text{low}}$(Hz) & $0.886_{-0.004}^{+0.004}\times 10^{-3}$ \cite{2004ragt.meet....1A} & 298$\pm$ 4 \cite{Kolos:2020ykz}  & 113 $\pm$ 5 \cite {Remillard:2006fc}  & 184$\pm$ 5 \cite{Remillard:2002cy} \\
     \\
  \hline\hline
\end{tabular} \label{freq}
\end{table*}

Markov Chain Monte Carlo (MCMC) analysis was performed using the Python library \texttt{emcee} \cite{2013PASP..125..306F} to determine constraints on the wormhole parameters in our study.
We employed the relativistic precession (RP) model in our analysis.

The posterior can be defined as in the reference \cite{Liu:2023vfh}:
\begin{equation}
\mathcal{P}(\Theta|\mathcal{D},\mathcal{M}) = \frac{P(\mathcal{D}|\Theta,\mathcal{M})\pi(\Theta|M)}{P(\mathcal{D}|\mathcal{M})},
\end{equation}
where $\pi(\Theta)$ is the prior and $P(\mathcal{D}|\Theta,\mathcal{M})$ is the likelihood. 
The priors are set to be Gaussian priors within boundaries, i.e., 
\begin{equation}
\pi(\Theta_i) \sim \exp\left(\frac{1}{2}\left(\frac{\Theta_i - \Theta_{0,i}}{\sigma_i}\right)^2\right),
\end{equation}
where $\Theta_{\text{low},i} < \Theta_i < \Theta_{\text{high},i}$ for the parameters $\Theta_i = [M, \alpha^*, \lambda, \ell^*]$ and $\sigma_i$ are their corresponding sigmas. 
Here we use the notation $\alpha^*=\alpha/M^2$ and $\ell^*=\ell_{3:2}/M$ is the normalized radial location of the 3:2 resonance. 
We take the prior values of the parameters of the wormhole, as presented in Table \ref{prior}.

In light of the upper and lower frequency data obtained in Section \ref{qposec}, our MCMC analysis is structured to utilize two distinct datasets. 
The central component of this analysis is the likelihood function, denoted as $\mathcal{L}$, and can be expressed as follows:

\begin{equation}
\log \mathcal{L} = \log \mathcal{L}_{\text{up}} + \log \mathcal{L}_{\text{low}},\label{likelihood}
\end{equation}
wherein $\log \mathcal{L}_{\text{up}}$ characterizes the likelihood associated with the upper-frequency data, given by:

\begin{equation}
\log \mathcal{L}_{\text{up}} = - \frac{1}{2} \sum_{i} \frac{(\nu_{\text{up, obs}}^i - \nu_{\text{up, th}}^i)^2}{(\sigma_{\text{up, obs}}^i)^2},
\end{equation}

On the other hand, $\log \mathcal{L}_{\text{low}}$ signifies the likelihood attributed to the lower frequency data, expressed as:

\begin{equation}
\log \mathcal{L}_{\text{low}} = -\frac{1}{2} \sum_{i} \frac{(\nu_{\text{low, obs}}^i - \nu_{\text{low, th}}^i)^2}{(\sigma_{\text{low, obs}}^i)^2},
\end{equation}

Here $i$ runs from 1 to an arbitrary integer, implying the number of measured upper and/or lower frequencies, $\nu^i_{\text{up, obs}}$ and $\nu^i_{\text{low, obs}}$ represent the observed results for the upper and lower frequencies, denoted as $\nu_{\text{up}}$ and $\nu_{\text{low}}$, respectively. 
Additionally, $\nu^i_{\text{up, th}}$ and $\nu^i_{\text{low, th}}$ correspond to their respective theoretical predictions. 
Furthermore, within these expressions, $\sigma^i_{x, \text{obs}^i}$ represents the statistical uncertainties associated with the given quantities.

\begin{table*}
\renewcommand\arraystretch{1.5} 
\caption{
The Gaussian prior of the wormhole in the beyond Horndeski  theory from QPOs for the selected X-ray sources. }\label{prior}
\begin{tabular}{lcccccccc}
\hline\hline
\multirow{2}{*}{Parameters} & \multicolumn{2}{c}{$SgrA^*$} & \multicolumn{2}{c}{GRO J1655-40}        & \multicolumn{2}{c}{GRS 1915+105} & \multicolumn{2}{c}{XTE J1550-564} \\
  & $\mu$ & \multicolumn{1}{c}{$\sigma$} & $\mu$          & $\sigma$         & $\mu$ & \multicolumn{1}{c}{$\sigma$} & $\mu$         & $\sigma$            \\
\hline
     $ M\; (M_{\odot})$ & $4.26\times 10^6$ & $0.12\times 10^6$ & 6.307& 0.066& 12.41& 0.62& 9.10&0.61 \\
     $\alpha^*$ & 1.20& 0.29& 1.20& 0.090& 1.20& 0.09& 1.20&0.12 \\
     $\lambda$ &0.50& 0.15& 0.45& 0.080& 0.20& 0.12& 0.23&0.05  \\
     $\ell^*$ & 6.07& 0.35& 5.68& 0.115& 4.84& 0.35& 4.95&0.35\\
     \hline\hline
\end{tabular}
\end{table*}

\begin{table*}
\renewcommand\arraystretch{1.5} 
\caption{\label{bestfitvalues}
     The best-fit values of the wormhole parameters from QPOs for the selected X-ray sources.}
\begin{tabular}{lcccc}
\hline\hline
Parameters  & $SgrA^*$ & GRO J1655-40   & GRS 1915+105 & XTE J1550-564 \\
\hline
     $ M\; (M_{\odot})$ & $4.25^{+0.12}_{-0.12}\times 10^6$ & $6.28^{+0.07}_{-0.06}$ & $12.19^{+0.53}_{-0.54}$ &  $9.02^{+0.21}_{-0.20}$ \\
     $\alpha^*$ & $1.18^{+0.29}_{-0.29}$ & $1.19^{+0.09}_{-0.09}$ & $1.21^{+0.09}_{-0.09}$ & $1.20^{+0.12}_{-0.12}$ \\
     $\lambda$ & $0.53^{+0.14}_{-0.14}$ & $0.57^{+0.07}_{-0.06}$ & $0.30^{+0.05}_{-0.04}$ & $0.30^{+0.03}_{-0.03}$ \\
     $\ell^*$ & $5.97^{+0.36}_{-0.36}$ &  $5.52^{+0.11}_{-0.11}$  &  $4.97^{+0.27}_{-0.28}$  &  $4.59^{+0.28}_{-0.28}$      \\
 \hline\hline
\end{tabular}
\end{table*}

\begin{figure*}
    \centering
     \includegraphics[width=0.48\textwidth]{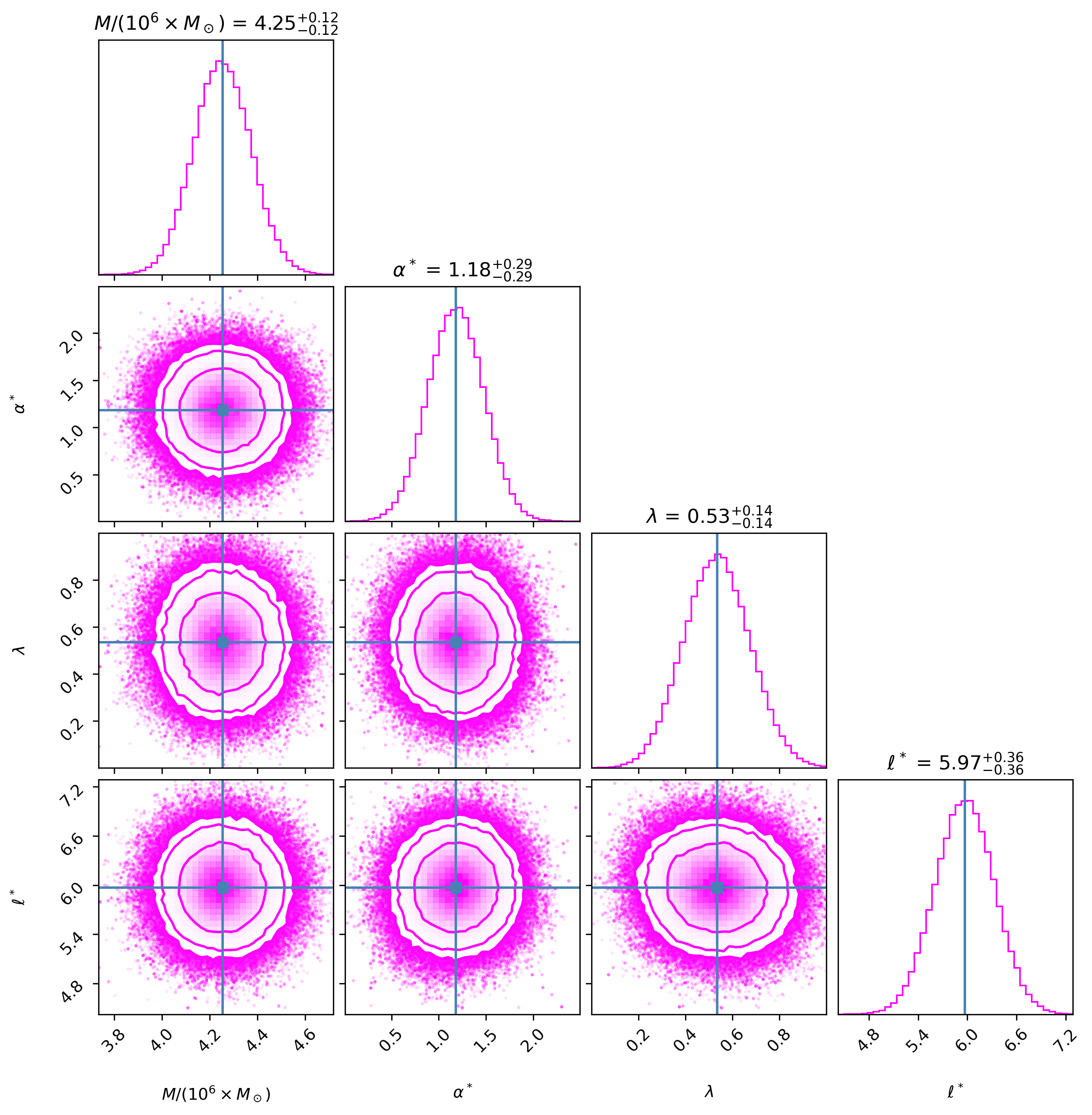}
    \includegraphics[width=0.48\textwidth]{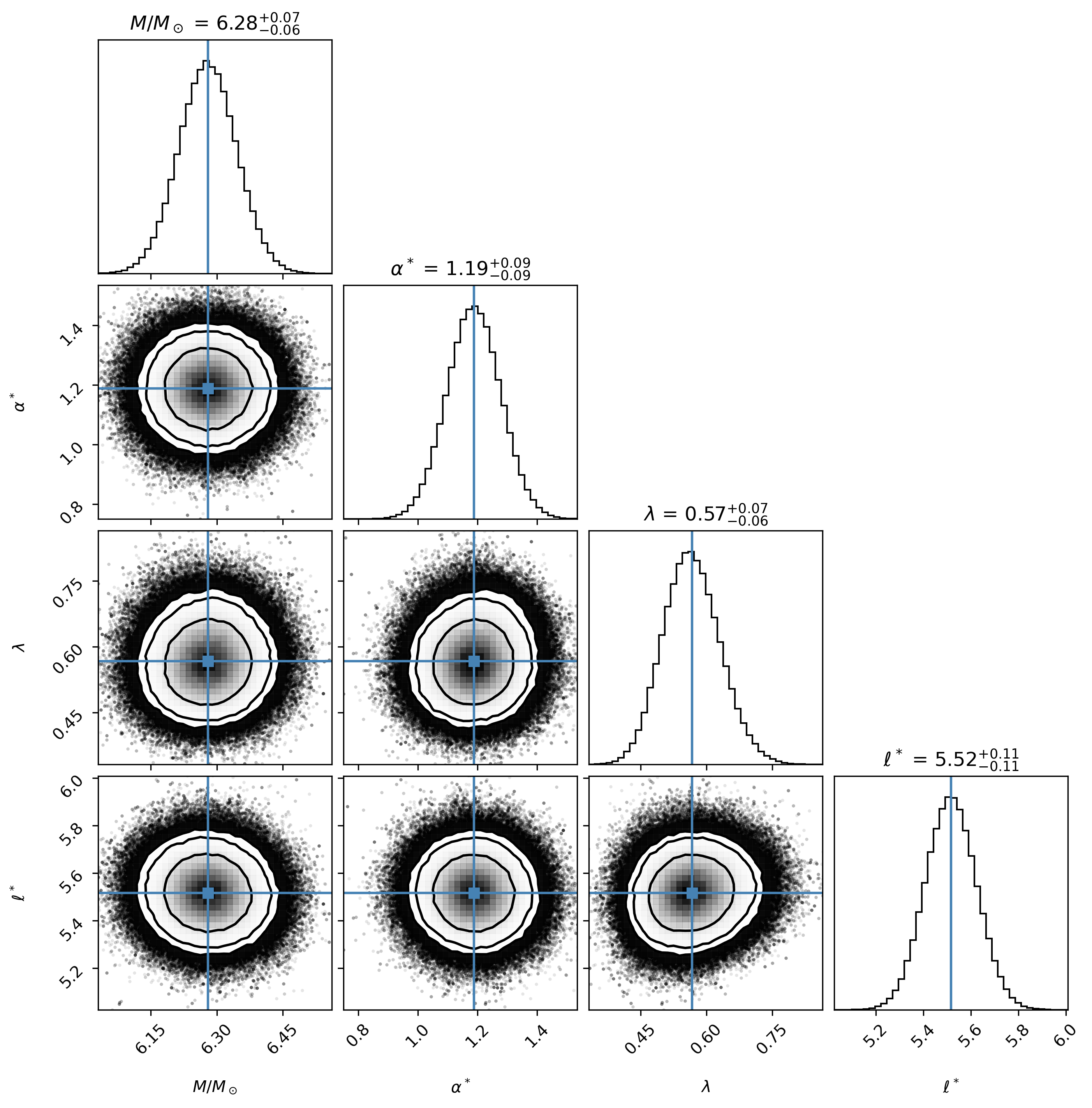}
    \includegraphics[width=0.45\textwidth]{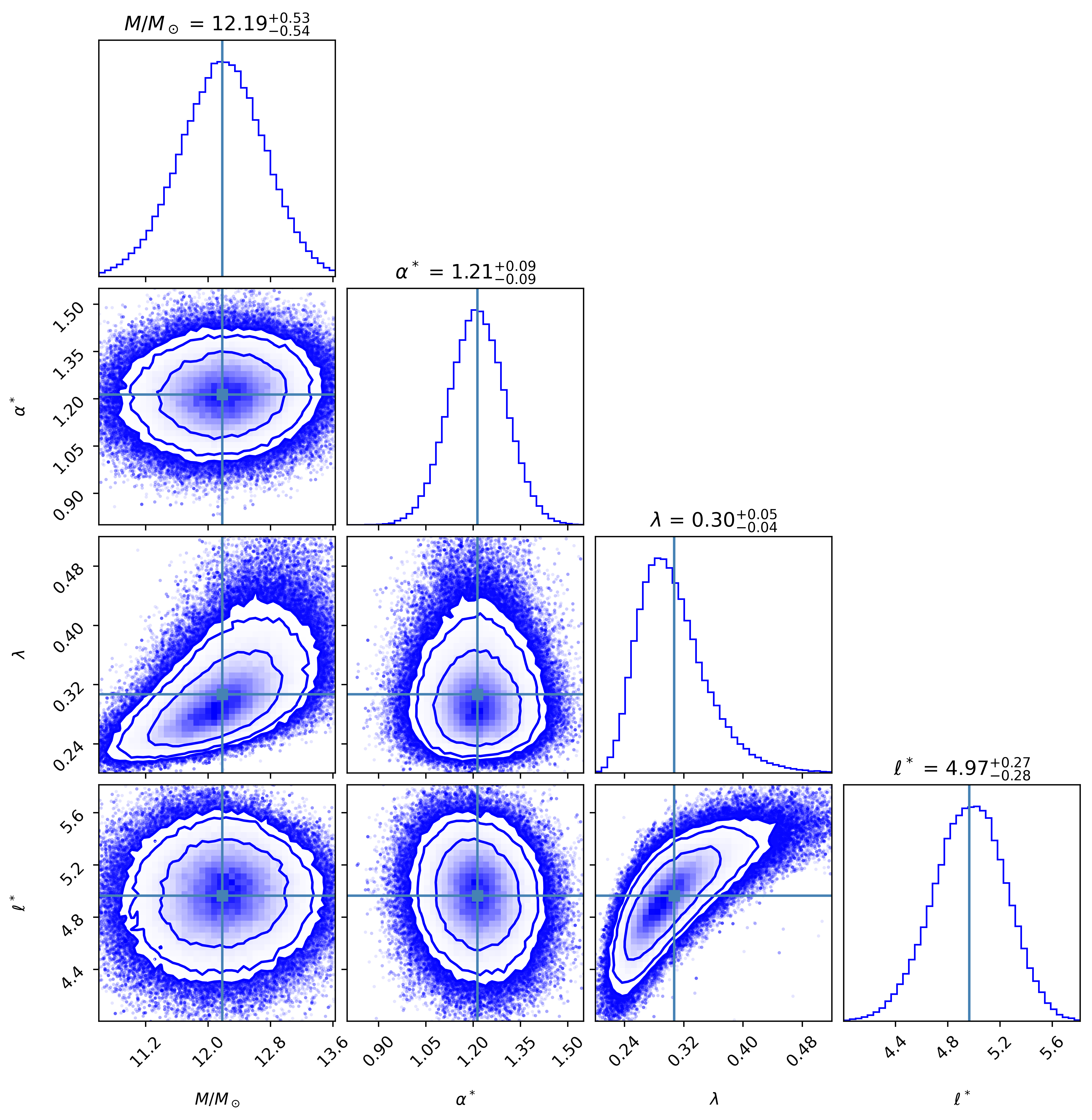}
    \includegraphics[width=0.45\textwidth]{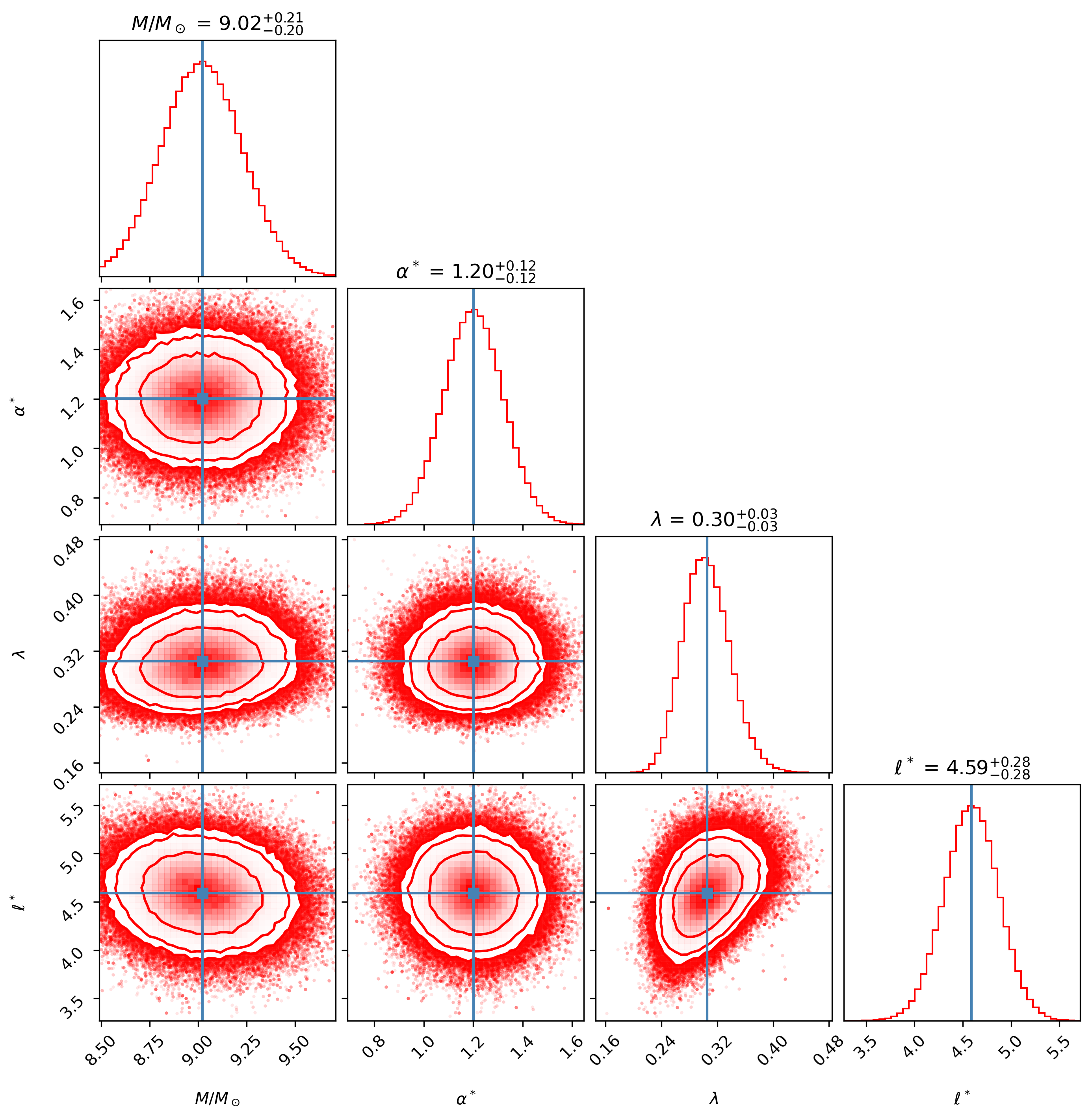}
    \caption{Constraints on wormhole mass, the $\alpha^*$, and $\lambda$ parameters in the $Sgr A^*$ (upper left), and microquasars GRO J1655-40 (upper left), GRS 1915+105 (lower left) and XTE 1550-564 (lower right) using MCMC code. Here $\alpha^*= \alpha/M^2$, and $\ell^*=\ell_{3:2}/M$ is the normalized radial location of the 3:2 resonance?}
    \label{fig:enter-label5}
\end{figure*}

\section{Conclusion: Results and Discussion}

Throughout this work, we have investigated the dynamics and oscillation of test particles together with the applications to QPOs observed near traversable wormholes in beyond Horndeski gravity theories.
The following main results are found:
\begin{itemize}
    \item  The wormhole throat decreases with the increase of the Horndeski parameters $\lambda$ and $\alpha$.
    \item  The angular momentum decreases slightly with an increase in both the parameters $\alpha$ and $\lambda$. However, at distances less than about, $\ell \simeq 4 M$ the angular momentum increases as $\lambda$ increases.
    \item An increase in $\alpha$ causes an increase in the effective potential. However, the gravitational effect of the parameter $\alpha$ disappears at $\ell>5M$. 
    \item Once we have $\alpha=0,2M$ \& $\lambda=0.1$, the ISCO of the particles coincides with the wormhole throat. As the parameter $\lambda$ increases, the ISCO goes far from the wormhole throat. 
    \item An increase in the $\lambda$ parameter causes a decrease in the upper and lower frequencies, and the QPO orbit with the frequency ratio 3:2, slightly shifts out.
    \item An increase of $\alpha$ reduces $\nu_U$ slightly near the wormhole throat (up to about $\ell \sim 3M$), and, at far distances (about $\ell\simeq 10M$), its effect is invisible.
    \item One can observe an increase in the highest values of the upper and lower frequencies corresponding to those generated at ISCO, due to an increase in $\alpha$.
    \item However, the highest value does not change in the variation of $\lambda$ and it only depends on $\alpha$, also the frequency ratio increases slightly with increasing $\lambda$.
    \item In the preceding section, we discussed the application of MCMC analysis for parameter estimation based on QPO data observed in the center of the microquasars GRO J1655-40, GRS 1915+105 \& XTE J1550-564 and Milky Way galaxy given in Table \ref{binary}.
    We have employed the MCMC code to explore the four-dimensional parameter space of the traversable wormhole solution. In this case, we have considered Sgr A* to be a supermassive wormhole candidate and studied three microquasars as stellar mass wormhole candidates. 
    The results of the MCMC analysis are presented in Fig. \ref{fig:enter-label5}. 
    The contours in these figures depict shaded regions representing the confidence levels of 1$\sigma$ (68\%), 2$\sigma$ (90\%), and 3$\sigma$ (95\%) for the posterior probability distributions of the complete set of parameters. 
    The figure comprises four panels: the upper-left panel corresponds to quasar Sgr $A*$, the upper-right panel to GRO J1655-40, the lower-left to GRS 1915 +105, and the lower-right to XTE 1550-564. 
    The details of prior values of the parameters $\mu$ and $\sigma$ which correspond to having the distribution of the values of the parameters $M$, $\alpha^*$, $\lambda$ and $\ell^*$ are given in Table \ref{prior} and their best-fit values for these objects are shown in Table \ref{bestfitvalues}. 
\end{itemize}

\section*{Acknowledgement}

This research is supported by Grant No. FA-F-2021-510 of the Uzbekistan Agency for Innovative Development. J.R., and B.A. acknowledge the ERASMUS + ICM project for supporting their stay at the Silesian University in Opava. 
J.K.~gratefully acknowledges support by the DFG project Ku612/18-1.
P.N.~ gratefully acknowledges support by the Bulgarian NSF Grant KP-06-H68/7.
F.A. gratefully acknowledges the support of Czech Science Foundation Grant (GA\v{C}R) No.~\mbox{23-07043S} and the internal grant of the Silesian University in Opava SGS/30/2023

\def\prc{Phys. Rev. C}
	\def\pre{Phys. Rev. E}
	\def\prd{Phys. Rev. D}
	\def\prl{Physical Review Letters}
	\def\jcap{Journal of Cosmology and Astroparticle Physics}
	\def\apss{Astrophysics and Space Science}
	\def\mnras{Monthly Notices of the Royal Astronomical Society}
	\def\apj{The Astrophysical Journal}
	\def\aap{Astronomy and Astrophysics}
	\def\actaa{Acta Astronomica}
	\def\pasj{Publications of the Astronomical Society of Japan}
	\def\apjl{Astrophysical Journal Letters}
	\def\pasa{Publications Astronomical Society of Australia}
    \def\pasp{Publications of the Astronomical Society of the Pacific}
	\def\nat{Nature}
	\def\physrep{Physics Reports}
	\def\araa{Annual Review of Astronomy and Astrophysics}
	\def\apjs{The Astrophysical Journal Supplement}
	\def\aapr{The Astronomy and Astrophysics Review}
	\def\procspie{Proceedings of the SPIE}
	
\bibliographystyle{apsrev4-1}
\bibliography{references,referenceJR,referencesJK,referencesPN}

\end{document}